# Water Markets as a Coping Mechanism for Climate-Induced Water Changes on the Canadian Economy: A Computable General Equilibrium Approach

Jorge A. Garcia and Roy Brouwer

**ABSTRACT**

Water markets represent a policy tool that aims at finding efficient water allocations among competing users by promoting reallocations from low-value to high-value uses. In Canada, water markets have been discussed and implemented at the provincial level; however, at the national level a study about the economic benefits of its implementation is still lacking. This paper fills this void by implementing a water market in Canada and examine how water endowment shocks would affect the economy under the assumptions of general equilibrium theory. Our results show a water market would damp the economic loss in case of reductions in water endowment, but it also cuts back on the economic expansion that would follow from an increase on it. These results provide new insights on the subject and will provide a novel look and reinvigorate informed discussions on the use of water markets in Canada as a potential tool to cope with climate-induced water supply changes.

## 1. INTRODUCTION
### 1.1 Overview

Water is essential for life and necessary for the well-being of the biosphere. Development and long-term sustainability of human societies imply having access to water in adequate quantities and acceptable quality. For this reason, foreseeable climate changes are a factor that must be pondered as they may alter the water cycle on a region and, consequently, bring about changes on availability, or even quality, of this resource.

Even though Canada is abundant in water resources, assessing the effect of potential changes in water availability on the economy is crucial for the design of adaptation or mitigation strategies by researchers or policy makers, and ultimately to inform society and spark behavioral changes if needed.

The interplay between water resources and economic impacts in Canada has been studied at the regional level for the Great Lakes (Garcia-Hernandez & Brouwer, 2020; Garcia-Hernandez, Brouwer, & Pinto, 2022) or the Saskatchewan river basin (Eamen, Brouwer, & Razavi, 2020). At the Canada-wide level, although water pricing has been studied before (Rivers & Groves, 2013), a comparison of the advantages of having a water market vs. current conditions, the effect of industry-wise water disruptions, or the asymmetric economic effects of water supply changes remain untested and unknown. This paper fills this void.

**1.2 Literature review**

Water market simulations have been fruitful for exploring market-efficient water allocations or gauging the adaptation of economies to water shocks. For example, Solís & Zhu (2015) use a CGE model to explore water markets as a way to cope with potential future water disruptions in a region of Spain. Koopman, Kuik, Tol, & Brouwer (2017) follow a similar approach for the Dutch economy where they test the inclusion of different industries into the water market trade. The effect of imposing water taxes on a water market and its overall consequences on the economy have been explored for South Africa (van Heerden, Blignaut, & Horridge, 2008). Overall, these studies show mixed results as benefits of implementing water markets (measured by welfare or GDP increase) depend on additional factors, such as which the industries are included in the trade or the allocation of the proceeds from water trading. The success of water markets also depends on the constraints imposed by infrastructure and transaction costs (McCann & Garrick, 2014)

There are few water-related CGE models developed for the Canadian economy. A similar model to the one presented here is found in Rivers & Groves (2013), where authors explore the welfare change in the Canadian economy if prices were used to allocate water resources to industries. Authors found either a welfare loss or gain depending on whether the proceedings of water payments go directly to households or to offset taxes, respectively. Other CGE models have also been used to study flooding events in (Gertz, Davies, & Black, 2019), where authors study the recovery response of Vancouver, British Columbia to cope with this natural event.

### 1.3 Novelty, relevance, and contributions

The model presented here, to the best of our knowledge, is the first CGE model constructed to study water markets in the Canadian economy as a whole. Although water markets have been discussed in Canada and even introduced before at the provincial level (Bjornlund, Zuo, Wheeler, & Xu, 2014), we hope the insights of this paper would provide a novel look on the subject and reinvigorate informed discussions. Besides the insights generated by the scenarios on the paper, the contributions of the present work are the development of a CGE model for Canada, as well as a procedure to create social accounting matrices from statistical data.

## 2. CGE MODEL

### 2.1 Overview

The CGE model closely follows the formulation presented in (Lofgren, Harris, & Robinson, 2002). The industry and commodity structure are shown in Figure 1. Raw water use is included as a primary input for the industries that belong to the general sectors of agriculture, mining, utilities and manufacturing (a total of 29). For the remaining industries, water use is assumed to be in the form of commodity. Figure A.2 shows the breakup of these industries.

The water sector (named "Water, sewage and other systems") and water-specific commodities (named "Water delivered by water works and irrigation" and "Sewage and dirty water disposal") are explicitly singled-out in the model. Irrigated and rainfed Irrigated crop production are treated as separate industries.

The current model assumes that water trade is not hampered by infrastructure constraints, and thus transaction costs and investment on water markets, though an important limiting factor, are not included in the present Canada-wide model since they are inherently dependent upon local conditions on infrastructure and water legislation.

Computationally, the CGE model is formulated as a square system (same number of variables as equations), where the user defines the values of the exogenous variables, which are the factor endowments, and the exogenous parameters, namely the elasticity substitutions of the industries and commodities. A solution to the CGE is an $x$ such that the following equations are met:

$$\left. \begin{array}{l} f(x, y \mid \theta_{exo}, \theta_{end}) = 0 \\ x \geq x_{min} \end{array} \right\} \quad (1)$$

where $f(\cdot)$ represents the system of equations of the CGE that encapsule the optimal behavior of economic agents; $x, x_{min}$ the model variables and their lower bound respectively; $y$ the exogenous (user-defined) variables; and $\theta_{exo}, \theta_{end}$ the exogenous and endogenous parameters of the model.

The calibration of parameters follows a two-step process: first, an optimization subroutine obtains the prices and quantities that match the values of the SAM (shown in Appendix A); next, these variables are used to calculate the endogenous parameters ($\theta_{end}$). Finally, the exogenous parameters (in line to those in literature) and variables (determined by user) are specified, and the simulation is executed to obtain $x$.

The model was implemented on GAMS® via its Python API and solved using the CONOPT® solver. Due to its extension, the model is presented in full on the Supplementary Information; however, the main aspects of the model are shown in the next sections.

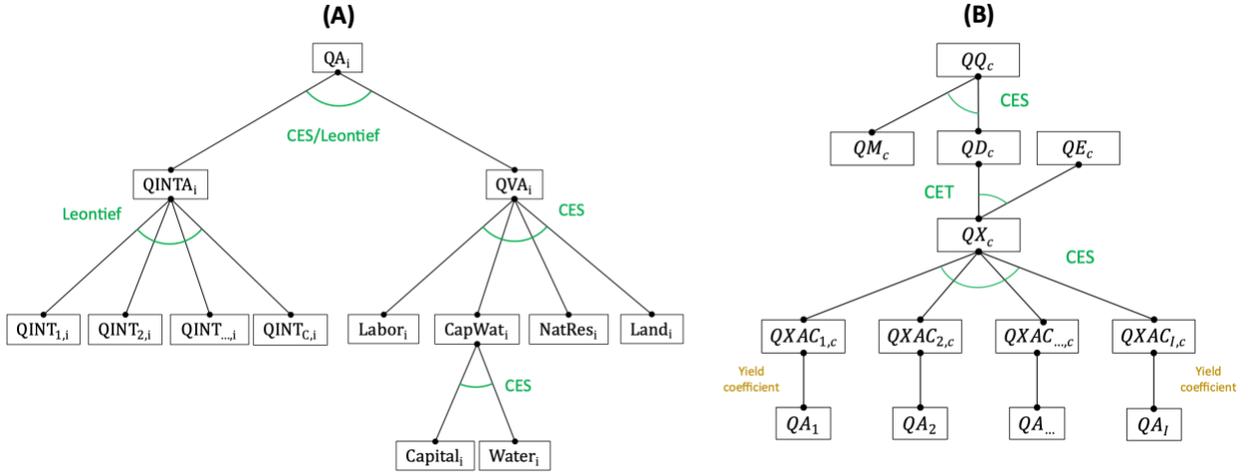

**Figure 1.** Industry (A) and commodity (B) production structure

## 2.2 Industry structure

The industry production structure has three levels. The top level consists of a constant elasticity of substitution (CES) or Leontief production function. Industries using a CES production function determine their production solving the following optimization problem which maximizes the gross revenue over production cost:

$$\max_{QVA_i,\ QINTA_i} PA_i\, QA_i - PVA_i\, QVA_i - PINTA_i\, QINTA_i, \qquad \forall i \in ICES \qquad (2.1)$$

s.t

$$QA_i = \alpha_i^I \left[ \delta_i^I\, QVA_i^{-\rho_i^I} + (1-\delta_i^I)\, QINTA_i^{-\rho_i^I} \right]^{-\frac{1}{\rho_i^I}}, \qquad \forall i \in ICES \qquad (2.2)$$

which yields the following first-order condition for optimality:

$$(1-\delta_i^I)\, PVA_i\, QVA_i^{1+\rho_i^I} = \delta_i^I\, PINTA_i\, QINTA_i^{1+\rho_i^I}, \qquad \forall i \in ICES \qquad (2.3)$$

where $ICES$ is the set of industries with a CES production function. $PA_i, QA_i$ are the price and quantity of industry $i$; $PVA_i, QVA_i$ the price and quantity of value-added, reflecting the cost of input factors; and $PINTA_i, QINTA_i$ the composite price and quantity of intermediate consumption purchased from other industries. Parameters $\alpha_i^I, \delta_i^I, \rho_i^I$ characterize the CES function; only the latter is user-defined and determines the elasticity of substitution. Only equations (2.2,2.3) are included into the model.

For industries using a Leontief production function, the associated equations, which are included in the model, are the following

$$\left.\begin{array}{l}QVA_i = iva_i\, QA_i \\ QINTA_i = inta_i\, QA_i\end{array}\right\} \quad \forall i \in ILEO \qquad (3)$$

where $ILEO$ is the set of industries with Leontief production function; $iva_i, inta_i$ are the shares of value-added and intermediate consumption with respect to the industry output.

The second level is a Leontief function for intermediate industry consumption and a CES function for value-added function at this level. Capital and water are treated as a single factor dubbed "CapWat"; the other factors are labor, land, and other natural resources. The value-added is determined by solving the following optimization problem, which maximizes value-added income over production costs:

$$\max_{QCW_i,\, QF_{f,i}|f\in FL1} PVA_i\, QVA_i - PCW_i\, QCW_i - \sum_{f\in FM\cap FL1} WFM_f\, QF_{f,i} - \sum_{f\in FNM\cap FL1} WFS_{f,i}\, QF_{f,i}, \quad \forall i \in I \quad (4.1)$$

s.t

$$QVA_i = \alpha_i^{va1}\left(\sum_{f\in FL1}\delta_{f,i}^{va1}\, QF_{f,i}^{-\rho_i^{va1}} + \delta_{CapWat,i}^{va1}\, QCW_i^{-\rho_i^{va1}}\right)^{-\frac{1}{\rho_i^{va1}}}, \quad \forall i \in I \qquad (4.2)$$

which yields the following first-order conditions for optimality:

$$(\alpha_i^{va1})^{\rho_i^{va1}} WFM_f\, QF_{f,i}^{(\rho_i^{va1}+1)} = PVA_i\, QVA_i^{(\rho_i^{va1}+1)}\, \delta_{f,i}^{va1}, \quad \forall i \in I, f \in \{FM \cap FL1\} \qquad (4.3)$$

$$(\alpha_i^{va1})^{\rho_i^{va1}} WFS_{f,i}\, QF_{f,i}^{(\rho_i^{va1}+1)} = PVA_i\, QVA_i^{(\rho_i^{va1}+1)}\delta_{f,i}^{va1}, \quad \forall i \in I, f \in \{FNM \cap FL1\} \qquad (4.4)$$

$$(\alpha_i^{va1})^{\rho_i^{va1}} PCW_i\, QCW_i^{(\rho_i^{va1}+1)} = PVA_i\, QVA_i^{(\rho_i^{va1}+1)} \delta_{CapWat,i}^{va1}, \qquad \forall i \in I \quad (4.5)$$

where FL1,FM,FNM are the sets of factors that belong to the first value-added function, mobile factors, and non-mobile factors; $PCW_i\, QCW_i$ the price and quantity of the composite CapWat; $QF_{f,i}$ the quantities of factor *f* allocated to industry *i*; $WFM_f, WFS_{f,i}$ the wages or price of the mobile and non-mobile factors. Parameters $\alpha_i^{va1}, \delta_{f,i}^{va1}, \delta_{CapWat,i}^{va1}, \rho_i^{va1}$ characterize the CES function. Only equations (4.2-4.5) are included in the model.

Finally, the third production level is a CES function describing the substitution between capital and water inputs. The associated optimization problem maximizes the income from the composite CapWat over its cost:

$$\max_{QCW_i,\ QF_{f,i}|f\in FL2} PCW_i\, QCW_i - \sum_{f\in FM\cap FL2} WFM_f\, QF_{f,i} - \sum_{f\in FNM\cap FL2} WFS_{f,i}\, QF_{f,i}, \qquad \forall i \in I \quad (5.1)$$

s.t

$$QCW_i = \alpha_i^{va2} \left( \sum_{f\in FL2} \delta_{f,i}^{va2}\, QF_{f,i}^{-\rho_i^{va2}} \right)^{-\frac{1}{\rho_i^{va2}}}, \qquad \forall i \in I. \quad (5.2)$$

The first-order conditions for optimality yield:

$$(\alpha_i^{va2})^{\rho_i^{va2}} WFM_f\, QF2_{f,i}^{(\rho_i^{va2}+1)} = PCW_i\, QCW_i^{(\rho_i^{va2}+1)} \delta_i^{va2}, \qquad \forall i \in I, f \in \{FM \cap FL2\} \quad (5.3)$$

$$(\alpha_i^{va2})^{\rho_i^{va2}} WFS_{f,i}\, QF2_{f,i}^{(\rho_i^{va2}+1)} = PCW_i\, QCW_i^{(\rho_i^{va2}+1)} \delta_i^{va2}, \qquad \forall i \in I, f \in \{FNM \cap FL2\} \quad (5.4)$$

where $FL2$ is the set of factors that belong to the second value-added function (namely, "capital" and "water"). Here again, the parameters $\alpha_i^{va1}, \delta_{f,i}^{va1}, \rho_i^{va1}$ determine the production function in (5.2). Only equations (5.2-5.4) are included in the model. The overall industry production structure is shown in Figure 1 (A). The elasticity of substitution among the primary factors in the first value-added function is set to 0.6, which is a value slightly lower than the one (0.7) in Solís & Zhu (2015) to reflect the lower substitution in a national model. The elasticity of substitution of the second level of the value-added function is set to 0.8, which is higher that the value used in Koopman et al. (2017) of 0.5, to reflect the fact that Canada is abundant on water resources and substitution between capital and water is less constrained.

The intermediate consumption in the second level of production is given by fixed technical coefficients as follows:

$$QINT_{c,i} = ica_{c,i} QINTA_i, \qquad \forall c \in C, i \in I \qquad (6)$$

where $QINT_{c,i}$ is the amount of commodity *c* consumed by industry *i*, and $ica_{c,i}$ is the technical coefficient expressing units of commodity *c* per unit of intermediate quantity *i*.

## 2.3 Commodity structure

Commodities are produced by industries using a fixed yield coefficient. Commodities of the same type are aggregated into a domestic supply following a CES function as follows

$$QX_c = \alpha_c^{ac} \left( \sum_{i \in I} \delta_{i,c}^{ac} \, QXAC_{i,c}^{-\rho_c^{ac}} \right)^{-\frac{1}{\rho_c^{ac}}}, \qquad \forall c \in CX \qquad (7.1)$$

$$\alpha_c^{ac(\rho_i^{ac})} \, PXAC_{i,c} \, QXAC_{i,c}^{(\rho_i^{ac}+1)} = PX_c \, QX_c^{(\rho_i^{ac}+1)} \delta_{i,c}^{ac}, \qquad \forall i \in I, c \in CX \qquad (7.2)$$

where $CX$ is the set of commodities with domestic output; $PX_c, QX_c$ the price and quantity of domestic output; $PXAC_{i,c}, QXAC_{i,c}$ those for the domestic output of commodity *c* by industry *i*. Parameters $\alpha_c^{ac}, \delta_{i,c}^{ac}, \rho_c^{ac}$ determine the composite domestic commodity output function. Equation (7.2) is the first-order condition for optimality.

The decision between allocating domestic supply to satisfy domestic or foreign consumption is determined using a constant elasticity of transformation (CET) function. Therefore, producers of domestic commodities determine the market to sell to based on:

$$QX_c = \alpha_c^t \left[ \delta_c^t\, QE_c^{\rho_c^t} + (1 - \delta_c^t)\, QD_c^{\rho_c^t} \right]^{\frac{1}{\rho_c^t}}, \quad \forall c \in (CD \cap CE) \quad (8.1)$$

$$\delta_c^t\, PDS_c\, QD_c^{(1-\rho_c^t)} = (1 - \delta_c^t)\, PE_c\, QE_c^{(1-\rho_c^t)}, \quad \forall c \in (CD \cap CE) \quad (8.2)$$

where CD and CE are the set of domestic commodities and exported commodities; $PDS_c, QD_c$ the domestic supply price and quantity of domestically produced and consumed commodities; $PE_c, QE_c$ the price and quantity of exports.

Domestic consumers decide between buying from local or foreign suppliers based on a CES function. This is implemented by the following equations:

$$QQ_c = \alpha_c^q \left[ \delta_c^q\, QM_c^{-\rho_c^q} + (1 - \delta_c^q)\, QD_c^{-\rho_c^q} \right]^{-\frac{1}{\rho_c^q}}, \quad \forall c \in (CD \cap CM) \quad (9.1)$$

$$(1 - \delta_c^q)\, PM_c\, QM_c^{1+\rho_c^q} = \delta_c^q\, PDD_c\, QD_c^{1+\rho_c^q}, \quad \forall c \in (CD \cap CM) \quad (9.2)$$

where $CM, PM_c, QM_c$ are the set, price, and quantity of imported commodities, respectively. Equation (9.2) is the first-order condition for optimality.

The commodity structure is shown in Figure 1 (B). The constant elasticities of substitution for commodities follow those used on the GTAP (T. W. Hertel, McDougall, Narayanan, & Aguiar, 1997).

## 2.4 Economic agents

There are four economic agents in the model: households (HH), non-profit institutions serving households (NPSH), corporations (CORP), and government (GOV). Each agent both a current and capital account. The use of agent capital accounts diverges from the model by (Lofgren et al., 2002) but follows the structure of income distribution used by the Canadian System of Macroeconomic Accounts.

The current income of agents is the sum of payments from the factors of production, taxes (for GOV only), and current transfers from other agents and the RoW. The government agent allocates transfers to other non-foreign agents based on the CPI while the other agents do this based on a fixed proportion of their total current income. Likewise, the RoW is assumed to transfer a fixed amount of foreign currency to domestic agents.

Expenditure is divided into consumption, current transfers to agents and RoW, and the remaining amount representing savings and investment is transferred to the respective agents' capital account. Consumption and transfers are fixed proportions of the total income of agents, therefore transfers to the capital accounts are endogenously determined by the model.

Disposable income is a fixed proportion of current income. For HH, the allocation of disposable income among commodities is divided into subsistence and marginal consumption. The subsistence consumption is a minimum commodity amount that must be met, and the marginal consumption is based on the remaining budget after all subsistence purchasing is met. Transfers are fixed proportions of income for non-government agents, and a fixed amount based on the CPI for GOV.

The capital accounts of agents receive income from their respective current account, capital transfers from other agents and RoW, and from domestic borrowing. These capital flows are allocated to gross fixed capital formation, change in inventories, capital transfers to agents and RoW, and domestic lending. Capital transfers among agents are also a fixed proportion of the capital income for non-government agents, and a constant amount based on the CPI for GOV. Gross fixed capital formation and inventories are kept in the same proportion as in the baseline for simplicity reasons.

Finally, financial flows are balanced by setting domestic borrowing plus RoW lending equal to domestic lending plus RoW borrowing. Closure of the capital accounts is set by letting domestic borrowing and RoW lending to be determined endogenously by the model. Domestic lending is set to keep the same proportion of capital income as that of the baseline for agents, and the same transfer with respect to the composite exchange rate (EXR) for RoW borrowing.

Since the CGE is a square system where the column summation equals the row summation, one row or column is redundant. Therefore, a dummy variable is created and set it equal to the financial flows balance equation, this is shown in the Supplementary Information.

## 3. DATA
### 3.1 Social Accounting Matrix

The development of the SAM for the Canadian economy follows a two-step procedure. First, a detailed SAM is constructed at the most detailed level available using different data sources from Statistics Canada.

The data are for the year 2018 and includes supply and use tables at the detail level (Statistics Canada, 2020h); current and capital accounts for households (HH) (Statistics Canada, 2020a, 2020b), non-profit institutions serving households (NPSH) (Statistics Canada, 2020c), corporations (Statistics Canada, 2020d, 2020e), general governments (Statistics Canada, 2020f), and non-residents (Statistics Canada, 2020g); and financial flows accounts (Statistics Canada, 2020i).

The resulting matrix contains 857 accounts in total, and follows a structure similar to the one presented in Mainar-Causapé, Ferrari, & McDonald (2018) and the three levels of income distribution for economic agents in Siddiqi & Salem (2012). The discrepancies between column- and row-wise summations were balanced using linear programming (see Appendix A).

The second step was to aggregate the detailed SAM and create the factor accounts for water intake, land, and other natural resources. The account aggregation follows the North America Industry Classification System (NAICS) level 2 (Statistics Canada, 2021a) for industries and the Input-Output Commodity Classification level 1 (Statistics Canada, 2019) for commodities. The water systems industry and water use commodities were explicitly singled out and kept separate. The creation of new factor accounts is shown in the next sections. The resulting accounts are shown in Table 1.

**Table 1.** SAM macro accounts

| No. | Set | Macro account | Accounts |
|---|---|---|---|
| 1 | C | Commodities | 64 |
| 2 | M | Transaction costs | 3 |
| 3 | I | Industries | 49 |
| 4 | T | Taxes | 3 |
| 5 | F | Factors | 5 |
| 6 | A | Economic agents | 4 |
| 7 | CAP | Capital accounts | 6 |
| 8 | FF | Financial flows | 1 |
| 9 | RoW | Rest of the world | 1 |
| | | Total | 135 |

**3.2 Water as Primary Input**

Water is included as primary input for industries that use water, i.e., industries belonging to agriculture, mining, power generation, water distribution, and manufacturing. The water use is taken from the physical water flow table (Statistics Canada, 2021d), which contains the water use by industry and year at the national level. Since the industry aggregation of the water use data is higher than that of the detailed SAM, water use is allocated to sub-industries based on their output. Payments for water use for mining, power generation and manufacturing are taken from Statistics Canada (2021c, 2021b). Since information on agriculture water payments is not publicly available, the mean value from the prices per meter cubic from the other sectors is taken to have a value that balances the low price paid by power generation and the relatively high price by manufacturing. The value obtained for agriculture (0.07 CAD/cubic meter) is within the bracket of agriculture water prices in the United States that goes from 0.005 to 0.1 [US/cubic meter] (Winchelns, 2010). These expenditures are taken as payments to water as a primary factor, which are subtracted from the payments to capital. The water rates are shown in Table 2 and the water user industries in Figure A2 in Appendix.

Table 2. Water intake rates [CAD/cubic meter]

| | |
|---|---|
| Power generation | 0.00581 |
| Mining | 0.05106 |
| Manufacturing | 0.16737 |
| Agriculture* | 0.07475 |

*estimated

These water intakes rates are taken as the initial price paid for water by the water-extracting industries that belong to those sectors.

### 3.3 Land and Other Natural Resources

Payments to land or other natural resources (exclusive of water) are calculated using the formula for industry-specific primary inputs from T. Hertel, et al. (2016), where the share of value-added for an industry-specific factor is given by

$$\theta_R = \frac{\theta_{VA} \, \sigma_{VA}}{\theta_{VA} \, \eta_S + \sigma_{VA}} \tag{10}$$

where $\theta_R, \theta_{VA}$ are the shares of resource and value-added respectively with respect to the total revenue of the industry, $\sigma_{VA}$ the elasticity of substitution of value-added, and $\eta_S$ the elasticity of supply of the resource factor. This step was performed on the detailed SAM.

### 3.4 Import Taxes and transaction costs

Import taxes and import transactions costs are not explicitly singled-out on the supply and use tables from the values paid by domestic commodities. Since import taxes most often vary depending on the province, commodity and quantity bought, it was decided to use as import tax rate and import transaction rate a value that is 1.5 times the rate paid by domestic commodities. This assumption avoids overcomplicating these calculations and makes the model sensible to the higher costs associated with imported commodities.

## 4. SCENARIOS

Scenario (A) simulates the economic costs on the Canadian economy of having targeted water shocks to selected water-intensive industries. Water input is industry-specific and thus not sharable among industries. Therefore, this scenario represents current conditions. The selected industries represent large water users:

- A1. Irrigated crop production.
- A2. Paper manufacturing.
- A3. Mining & quarrying.
- A4. Water sectors.
- A5. Power generation.

Scenario (B) assumes a water market is implemented as the mechanism to allocate water intake to economic activities and explores the response on the Canadian economy of water supply shocks due to potential climate changes. Therefore, water input is mobile across industries in this scenario.

In both scenarios labor is assumed mobile, whereas and capital, land, and other natural resources are industry-specific.

## 5. RESULTS

Experiments of scenario A (Figure 2) show that increasing water scarcity on the targeted industries decreases the output except for Mining & quarrying. Output is most sensible to shocks in Paper manufacturing, Power generation, and Irrigated crop production. In terms of GDP, shocks to Paper manufacturing and Irrigated crop production produce the largest changes. Water scarcity also have consequences to the Canadian balance of trade where shocks to Paper manufacturing increases the most the baseline deficit and those to the water sector have the least impact.

Overall, water shocks to Paper manufacturing, Irrigated crop production, and Power generation have the most sizeable impacts on the Canadian economy.

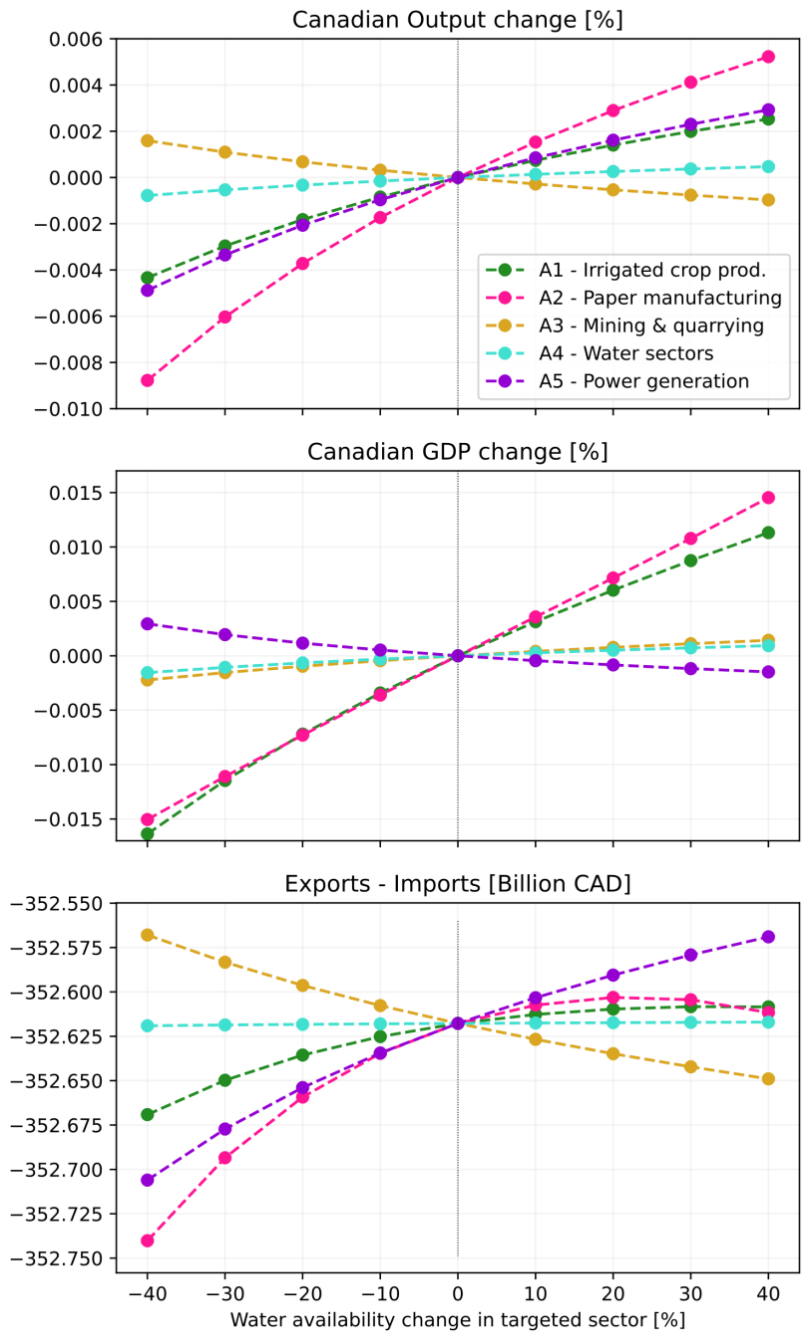

**Figure 2.** Results for scenario A

Results from scenario B, where a water market is implemented, show the Canadian economy would have a relatively small response to water supply changes. The largest affected industry is primary metal manufacturing with an output change in the range of [-3.7%, 2.2%], which is a small variation compared to the change in water availability. The total output of the economy has a small response between [-0.1%, 0.05%] (see Figure 3), while that of GDP is between [-0.02%,0.04%]. Results in the same order of magnitude have been found before for the Canadian economy (Rivers & Groves, 2013).

Another insight from this scenario is the linkage between water availability and trade balance, which indicates that when water becomes scarce and its price increases, the production of domestic water-intensive commodities becomes costlier. These commodities then must be satisfied on foreign markets, which amplifies the trade deficit on the economy. This linkage also works in the opposite direction when water supply is increased, though the effect is less pronounced as seen in Figure 3.

In terms of income, that of HH shows a positive correlation with water supply (Figure A3 in Appendix). This is because a decrease in water supply, forces industries to employ more labor, which in turn makes labor less productive and have a lower price. The effect is the opposite when water supply is increased.

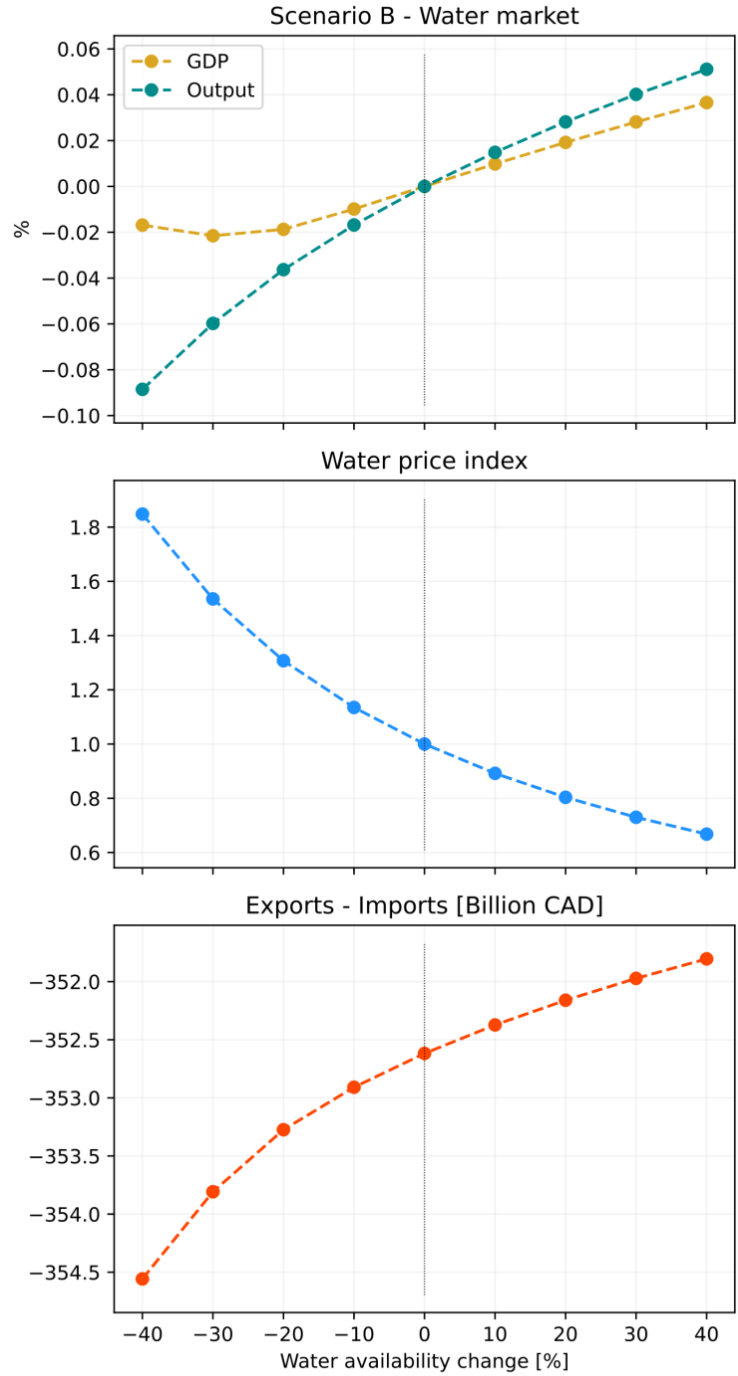

**Figure 3.** Results for scenario B

The income of corporations (Figure A3 in Appendix) shows a negative correlation with water supply. This is likely due to capital being non-mobile and associated with water in the capital-water production function. Therefore, a variation water input is only met with capital price changes (since capital supply is fixed) which has the effect that a decrease in water supply increases the relative contribution of capital to production, which overall increases capital income. Since capital income benefits primarily to corporations, hence the inverse relation.

The income of governments and NPSH exhibits a more complex behavior, mainly due to both being the recipients of large income transfers from HH and corporations (jointly accounting for 27% and 38% of their income respectively) that affect them in opposite direction as stated above.

Overall, water scarcity in the scenario where a water market is implemented (B) produces a redistribution of income affecting households, nonprofit institutions serving households, and the general government and benefiting corporations (Figure A3 in Appendix).

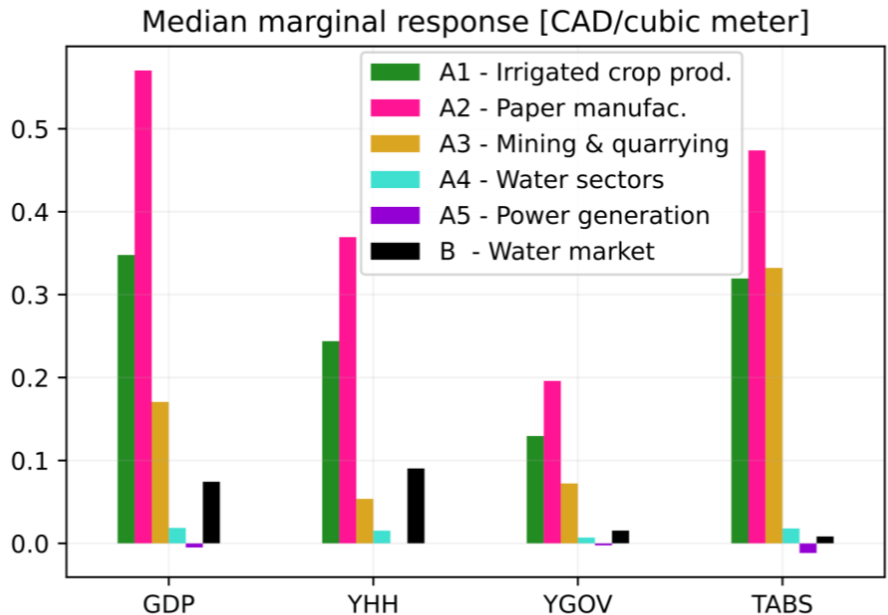

**Figure 4.** Comparison of median marginal response across the domain ($\Delta w \in [-40\%, 40\%]$) of water shocks for the scenarios tested; YHH: Canadian households income; YGOV: Canadian government income; TABS: total absorption, total domestic consumption at market prices

In order to make the results from scenarios A and B comparable, the median marginal response on four variables is calculated for each simulation experiment: GDP, households income (YHH), government income (YGOV), and total absorption (TABS) which is the market value of the domestic consumption.

The largest marginal response (CAD/cubic meter), shown in Figure 4, is produced by shocks to paper manufacturing, followed by those to Irrigated crop production. Power generation and the water sector produce small economic responses per unit of volume of water change. The scenario where the water market is implemented, seem to balance out the economic response observed on the industry-specific shocks on GDP, YHH and YGOV. This damping effect acts on both directions: reducing the economic loss when water endowment is decreased, but also capping the economic gain when water endowment is increased. Domestic consumption, as measured by TABS, is mostly unaffected under scenario B unlike the results from scenario A.

In terms of the distribution of impacts across industries, the Gini index is calculated over the absolute relative output change[1] as a measure of the concentration of industry output changes. This index shows that the scenario with the most uniform output effect corresponds to that of the water market, whereas the scenario with the narrowest effect or unequal distribution is where the water sector is affected. Table 3 shows the median of the Gini index calculated for each water shock level on each scenario.

**Table 3.** Median Gini index on output water shocks

| Scenarios | | | | | |
|---|---|---|---|---|---|
| **A1** | **A2** | **A3** | **A4** | **A5** | **B** |
| 0.79 | 0.89 | 0.76 | 0.93 | 0.73 | 0.73 |

---

[1] Absolute relative output change calculated as $abs(QA_i^{(new)} - QA_i^{(baseline)})/QA_i^{(baseline)}$ for each industry $i$.

## 6. SENSITIVITY ANALYSIS

### 6.1 Change in rainfed output

A major source of uncertainty in the data is found on the percentage of Irrigated crop production output that comes from rainfed (non-irrigated) areas. Data on irrigated farms from Statistics Canada (Statistics Canada, 2021e) indicate that about 80% of farms are irrigated (7,015/8,430 ≈ 0.83) for 2018; whereas the World Bank sets to 1.6% the agriculture irrigated land as percentage of total agricultural land (World Bank, 2015). A conservative value was selected for the baseline scenario, assuming that 20% of the crop output comes from irrigated land. Nonetheless, this estimate may be not be accurate because it does not consider the size of the farm or the market value of the crops. For this reason a sensitivity analysis is performed on the percentage of rainfed Irrigated crop production (Figure 5).

The results show that although increasing the percentage of irrigated output increases the costs of water shocks, its effect on the Canadian GDP is insignificant (less than 20 million CAD per year). There is also a clear trend showing that as the water shock increases, the difference in cost with respect to baseline increases as well. This pattern is only interrupted for a -40% shock, which shows that at this level of reduction the difference in GDP loss is minimal. Therefore, the results of scenario B are mostly unaffected if the percentage of rainfed crops is in the range of [20%,80%].

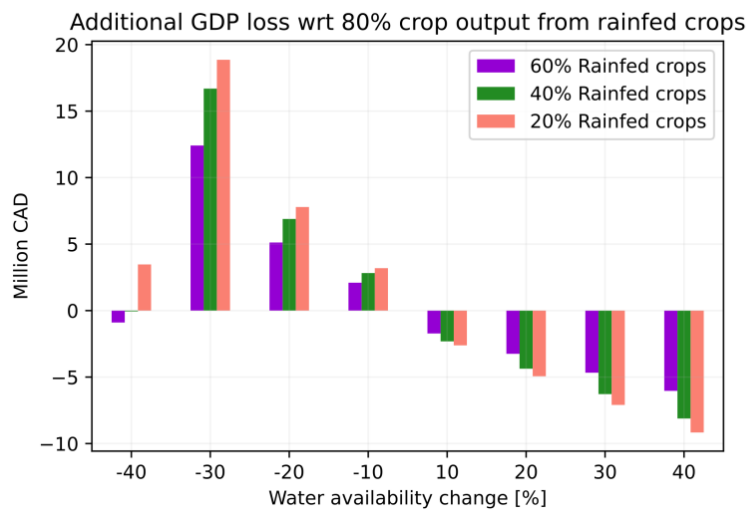

**Figure 5.** Cost in GDP with respect to the baseline scenario B

## 6.2 Change in water input valuation

A second major source of uncertainty relates to the payments to extract water for the industries that use this resource as primary factor. To assess the effect of having higher water costs, and therefore higher water prices, three additional cases are simulated for scenario B where the valuation of the water input is two, three, and four times the value of the baseline scenario.

**Table 4.** Median marginal response wrt baseline
[additional CAD/cubic meter]

|     | 2x valuation | 3x valuation | 4x valuation |
| --- | --- | --- | --- |
| GDP | 0.06 | 0.11 | 0.15 |
| YHH | 0.07 | 0.14 | 0.12 |

Results in Table 4 show that increasing water input costs has the effect of increasing the sensibility of GDP to water variations. For a factor of two increase in water input valuation, the GDP response is about twice as much as in the baseline scenario, but this factor decreases for higher valuations. HH income increases with water valuation for the first two cases, but it slightly decreases for the last case. Overall, increasing the cost of water input by a factor $n$, has a response on GDP and HH income, roughly speaking, less o equal to $n$ times the original value for the water shocks range tested.

## 7. DISCUSSION & CONCLUSIONS

This paper presented a water computable general equilibrium (CGE) model for the Canadian economy which aims at exploring the effect of implementing a water market as a tool to cope with water level changes. The development of the CGE model entailed the creation of a social accounting matrix (SAM) from statistical data. Physical water flows and capital payments to obtain water by sectors are used to create the water price rates.

A novel insight produced by this study is that the Canadian economy responds more pronouncedly to shortages than exceedances on water endowment regardless of whether a water market is in place or not. It was also found that water variations have a relatively small effect overall on the economy, which goes in agreement with previous results in the literature (Rivers & Groves, 2013).

Implementing a water market across Canada has the effect of balancing out the economic costs that otherwise industry-specific water disruptions would produce. This effect acts in both directions, damping the economic loss due to water cutbacks but also the gains from increasing the water endowment.

If a water market is implemented, water variations affect mostly, in relative sense, to the primary metal manufacturing and paper manufacturing industries, followed by Irrigated crop production and the Water sectors.

Overall, the implementation of water markets deserves a closer look to incorporate the effect of spatially differentiated disruptions and the transaction costs. We believe these two aspects are better addressed on a spatially disaggregated model that would allow to introduce several local water markets. This remains a future direction to expand the CGE model of the Canadian economy.

## ACKNOWLEDGEMENTS

This work was financially supported by the project Global Water Futures, funded by the Canada First Research Excellence Fund (CFREF), in particular the sub-projects Integrated Modelling Program for Canada (IMPC) and Lakes Futures.

https://data.worldbank.org/

# Appendix A. Balancing of detailed SAM

The optimization subroutine to balance the detailed SAM is the following:

1. Construct the detailed SAM from the data sources and calculate the difference between the column- and row-wise summation for each account.
2. Identify the accounts whose absolute difference > tolerance (1E-5).
3. Ensure the sum of differences is equal to zero, i.e., the errors cancel each other out.
4. Construct a submatrix that has as rows and columns the accounts with sum discrepancies. In all years calculated, these accounts belonged to the economic agents (A) and capital accounts (CAP).
5. Contruct a linear program to find the transfers among these accounts that balance all the accounts. These adjustments are made on the current and capital transfers matrices among agents and RoW, which are those at the intersections (A, A), (CAP, A), (CAP, CAP), (CAP, ROW), (ROW, CAP) seen in Figure A1.

| | C | M | I | T | F | A | CAP | FF | RoW |
|---|---|---|---|---|---|---|---|---|---|
| C | | Margins consumption | Intermediate consumption | | | Final consumption | Investment and stock changes | | Exports |
| M | Transaction margins | | | | | | | | |
| I | Domestic production | | | | | | | | |
| T | Net commodity taxes | | Net Industry taxes | | | | | | |
| F | | | Payments to factors | | | | | | |
| A | | | | Tax revenue | Income agents | Transfers and income taxes | | | Current transfers from RoW |
| CAP | | | | | | Depreciation and savings | Capital transfers | Borrowing | Capital transfers from RoW |
| FF | | | | | | | Lending | | Borrowing from RoW |
| RoW | Imports | | | | | Current transfers to RoW | Capital transfers to RoW | Lending to RoW | |

**Figure A1.** Depiction of the aggregated SAM

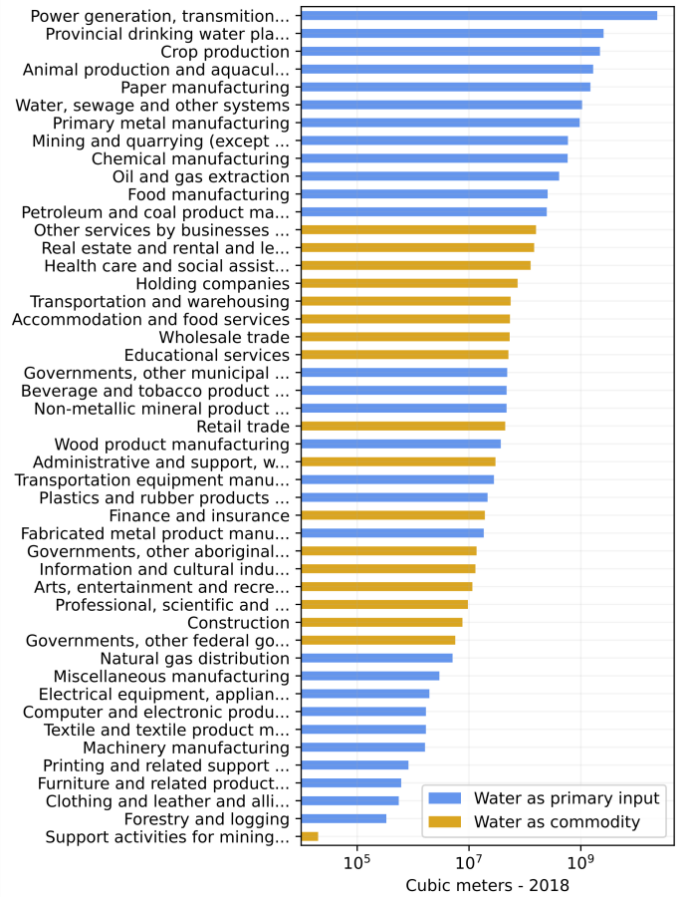

**Figure A2.** Water use by industry (only industries with reported water use are shown)

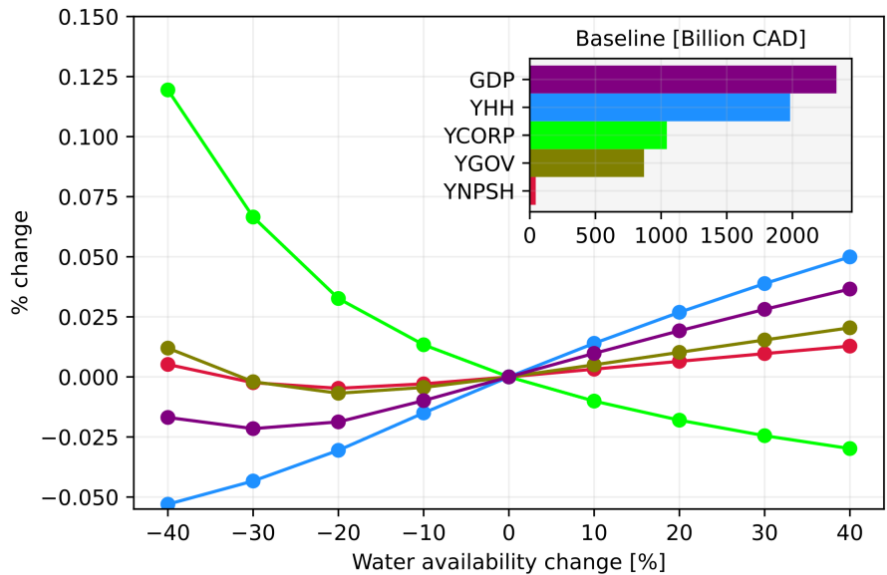

**Figure A3.** GDP and agents' income change for scenario B

28